\RequirePackage{etex}
\documentclass{article}
\usepackage{ijcai21}

\usepackage{times}
\usepackage{soul}
\usepackage{url}
\usepackage[hidelinks]{hyperref}
\usepackage[utf8]{inputenc}
\usepackage[small]{caption}
\usepackage{graphicx}
\usepackage{amsmath}
\usepackage{amsthm}
\usepackage{booktabs}
\usepackage{algorithm}
\usepackage{algorithmic}
\usepackage{graphicx}
\usepackage{tikz}
\usepackage{tikz-network}
\usepackage{caption}
\usepackage{subcaption}
\usetikzlibrary{shapes}
\usepackage{dblfloatfix}

\newtheorem{theorem}{Theorem}

\title{Emerging Methods of Auction Design in Social Networks}

\author{
Yuhang Guo
\And
Dong Hao\footnote{corresponding author.}
\affiliations
University of Electronic Science and Technology of China\\

\emails
guoyuhang@std.uestc.edu.cn,
haodong@uestc.edu.cn.
}

\begin{document}

\maketitle

\begin{abstract}
In recent years, a new branch of auction models called diffusion auction has extended the traditional auction into social network scenarios. The diffusion auction models the auction as a networked market whose nodes are potential customers and whose edges are the relations between these customers. The diffusion auction mechanism can incentivize buyers to not only submit a truthful bid, but also further invite their surrounding neighbors to participate into the auction. It can convene more participants than traditional auction mechanisms, which leads to better optimizations of different key aspects, such as social welfare, seller's revenue, amount of redistributed money and so on. The diffusion auctions have recently attracted a discrete interest in the algorithmic game theory and market design communities. This survey summarizes the current progress of diffusion auctions.
\end{abstract}
\section{Introduction}
Auction design is a common paradigm and a successful application of market design. Now it has become a representative interface integrating economics and artificial intelligence. Convening a larger group of customers  in an auction may possibly increase not only the seller's revenue but also the social welfare. However, classic auction models focus on implementing desired social choices in a fixed group of bidders who can be directly informed by the seller. They do not take the underlying social network among the agents into account. On the contrary, in any real markets, the social and economic relations between entities play an important role. 

The role of social networks in auctions is twofold. First, an agent's preference for the selling item is inherently associated with her position in the social network. More importantly, it is the social networks where the sale information spreads and where the commodities flow. That is, the crowds and markets are networked. Without careful consideration of social networks, the sale may be blocked within a local community, leaving many high valuation buyers excluded from the sale. As a result, the auctioneer's revenue and the social welfare can only be locally optimized. 

The main difficulty for the auction mechanism to convene more buyers 
lies in the conflict that although the seller wishes to attract more people to join the auction in order to increase her revenue, the buyers have no incentive to bring more competitors into the auction. This essentially reflects the conflict between the system’s optimality and the individuals' self-interests. 
To tackle this general problem, in recent years, the diffusion auction has been proposed. 
Diffusion auction consists of two components: agents' spontaneous expansion of the market and the seller's implementation of allocation and pricing in the expanding market. Since diffusion auction design can convene more participants than traditional mechanism design, it can simultaneously improve the social welfare and the seller's profit, which is known to be a difficult objective. Furthermore, with the development of online social network technology, the interaction between agents is becoming easier, faster and broader, making diffusion auction design more realistic.
This survey gives a comprehensive survey of current progress of diffusion auction design.
\begin{figure}[!htbp]
    \centering
    \includegraphics[width=0.47\textwidth]{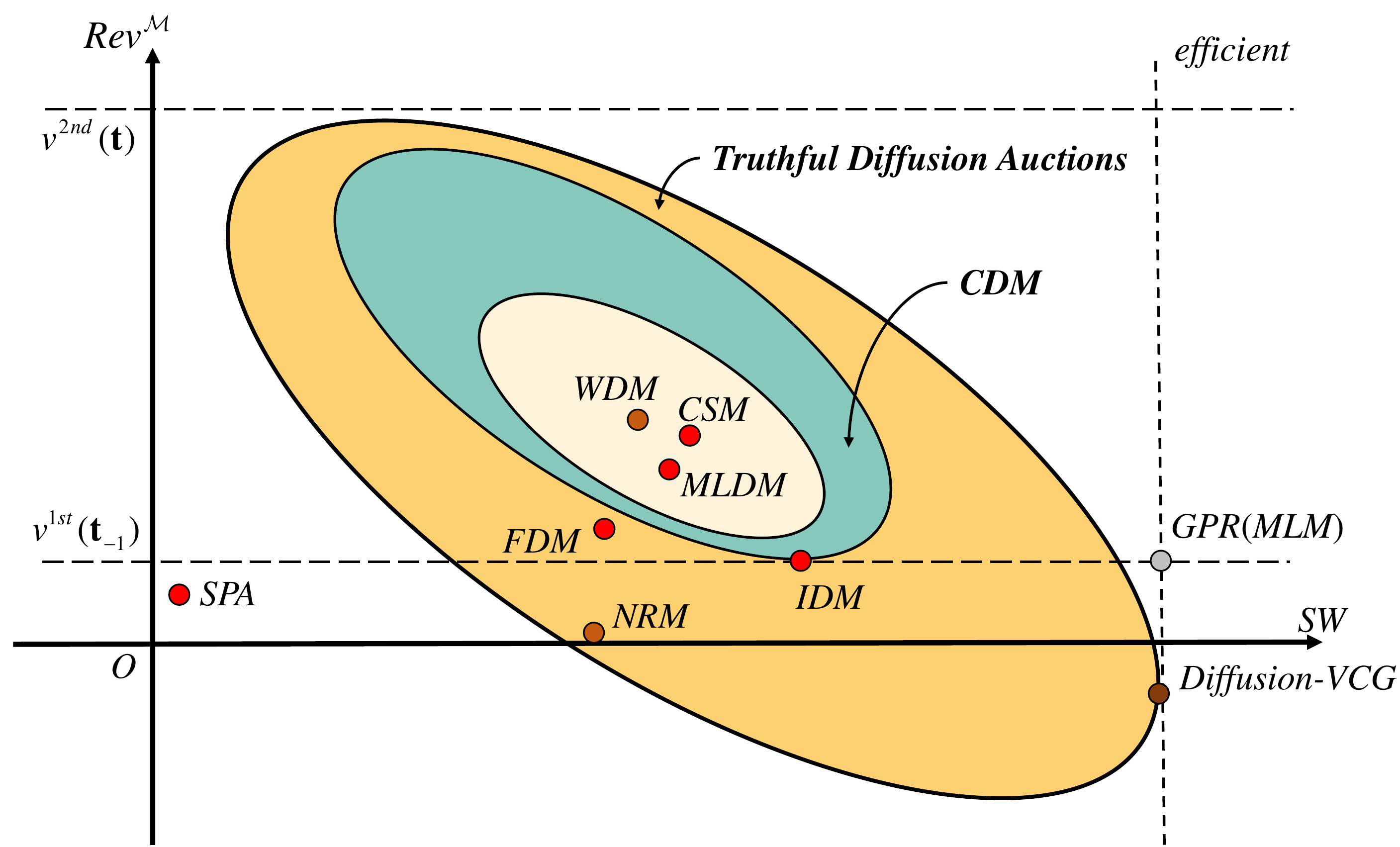}
    \caption{Research map of representative diffusion auction mechanisms under a certain social network. The $x$-axis is the value of social welfare and the $y$-axis is seller's revenue. Each circle is a certain class of diffusion auction mechanisms. Each dot is a specific auction mechanism.}
    \label{fig_mechanisms}
\end{figure}

For auction design, efficiency and the seller's revenue are two central criteria, but usually they conflict with each other. With the new concepts and models, various novel diffusion auction mechanisms have aligned the optimization of these two criteria in different ways. We illustrate the research map of the most representative mechanisms of diffusion auction in Figure \ref{fig_mechanisms}. The traditional second-price auction (SPA) sells the item in a local market with relatively low efficiency and revenue. The pioneer works of the diffusion auction \cite{li2017mechanism,lee2017mechanisms} have proved that classic VCG mechanism \cite{vickrey1961counterspeculation,clarke1971multipart,groves1973incentives} can be naturally extended to social network scenarios but is not budget balanced. Especially, Li \emph{et al.} \shortcite{li2017mechanism} showed that VCG could incur the seller with a large deficit in extreme network cases. Lee \shortcite{lee2017mechanisms} analysed the conditions for VCG's deficit. Based on these observations of VCG's shortcoming in social networks, Li \emph{et al.} \shortcite{li2017mechanism} further proposed a formal method IDM for modeling the information diffusion and designed the very first incentive-compatible diffusion auction. Lee \shortcite{lee2017mechanisms} proposed another diffusion mechanism MLM which is efficient, has equal revenue with IDM but can not elicit truthfulness. The yellow region in Figure \ref{fig_mechanisms} circles the domain of all strategyproof diffusion auctions, with the diffusion version of VCG on a border where the efficiency is maximized but the revenue could even be negative. After these two initial works, CSM \cite{li2018customer} extended IDM into economic networks. Then, CDM and WDM \cite{li2019diffusion} formulated solutions for unweighted and weighted networks respectively. Also, FDM \cite{zhang2020redistribution} and NRM \cite{zhangECAIincentivize} studied redistribution issues via social networks. Besides these diffusion auction models, Influence Maximization (IM) is a large branch of works investigating how to find influential nodes and how to maximize information propagation \cite{kempe2003maximizing,chen2009efficient,banerjee2019maximizing}. However, for auction in social networks, an agent needs to reason about her strategy which is a couple of bid and diffusion actions. The IM methods are not applicable for this complex market.

Section 2 explains the fundamental concepts and model of diffusion auctions. Section 3 clarifies and compares diffusion auction mechanisms under different scenarios. Section 4 further introduces incentive design for spontaneous information diffusion under various scenarios other than auction. Section 5 concludes this survey and shows future directions.

\section{Modeling of Diffusion Auction}
\subsection{Networked Auction and Information Diffusion}
A social network can be regarded as a directed graph $G=(N,E)$, in which the nodes $N$ represent agents and the edges $E$ represent the social links between agents. Let $N=\{1,\dots,n\}$ denote the agent set. For each $i\in N$, there is a set of neighbors $r_i \subseteq N \setminus \{i\}$. Agent $i$ can directly communicate with her neighbors, but cannot communicate with other agents. Assume in this social network, there is a seller $s$ who has items to sell and all other agents $i \in N \setminus \{s\}$ are potential buyers of this item. At the very beginning, only the seller's neighbors know the sale and the auction is run only among this local community. This is the general assumption of classic auction theory. On the contrary, the diffusion auction considers a more realistic scenario where after knowing the sale, agents may possibly share this sale information with their neighbors. 

Assume $v_i$ is $i$'s valuation of the item and $r_i \in {\mathcal P}(r_i)$ is her neighbor set where ${\mathcal P}(r_i)$ is the power set. Agent's type or action is defined by a tuple $t_i=(v_i,r_i)$ and the type profile is ${\mathbf{t}}=(t_1,\cdots,t_n)$. The type profile except agent $i$ is $\mathbf{t}_{-i}=\{t_1,\cdots,t_{i-1},t_{i+1},\cdots,t_n\}$. Thus the type of each agent has been expanded from classic single dimensional valuation to a non-typical multi-dimensional structure where one dimension is the valuation and the other dimensions are neighbors. 
The type space is ${T}_i:V_i\times {\mathcal P}(r_i)$ and the type profile space is $T=\times T_{i}$. 
For any two different types $t_i^1=(v_i^1,r_i^1)$ and $t_i^2=(v_i^2,r_i^2)$, the diffusion auctions assume $t_i^1\succ t_i^2$ holds \textit{iff} $v_i^1\geq v_i^2 \wedge r_i^1\subseteq r_i^2$, i.e. reporting a higher bid and diffusing information to fewer neighbors give a bidder a higher probability to win. We say a type profile $\mathbf{t}$ from a subset of agents is \emph{feasible} when there is a maximum connected subgraph containing the seller and these agents. For example, if all the $n$ agents in the social network know the auction, ${\mathbf{t}}=(t_1,\cdots, t_n)$ is feasible.

A diffusion auction mechanism $\mathcal{M}=(\boldsymbol{\pi},\mathbf{x})$ consists of an allocation rule $\boldsymbol{\pi}=\{\pi_i\}_{i\in N_{-s}}$ and a payment rule $\mathbf{x}=\{x_i\}_{i\in N_{-s}}$. Similar to conventional auction settings, let $t_i'=(v_i',r_i')$ be bidder $i$'s reported type where $r_i'\subseteq r_i$ means $i$ diffuses the sale information to a subset of her neighbors $r_i'$. Given an allocation policy $\boldsymbol{\pi}$ and buyers' reported type profile $\mathbf{t}'$ , the \emph{social welfare} of an outcome is defined as $SW(\boldsymbol{\pi},\mathbf{t}')=\sum_{i\in N_{-s}}\pi_i(\mathbf{t}') v_i$. Given bidder $i$'s true type $t_i=(v_i,r_i)$ and reported type profile $(t_i',\mathbf{t}_{-i}')$ and a mechanism $\mathcal{M}=(\boldsymbol{\pi},\mathbf{x})$, we define bidder $i$'s quasi-linear \emph{utility} as $u_i(t_i,(t_i',\mathbf{t}_{-i}'),(\boldsymbol{\pi},\mathbf{x}))=\pi_i(t_i',\mathbf{t}_{-i}')v_i - x_i(t_i',\mathbf{t}_{-i}')$. From the prospective of seller $s$, her utility is defined as $Rev^{\mathcal{M}}(\mathbf{t}')=\sum_{i\in N_{-s}}x_i(t_i',\mathbf{t}_{-i}')$.

Next, some concepts used to assess the properties of diffusion mechanisms will be given. First of all, \textit{efficiency}: Given a diffusion mechanism $\mathcal{M}=(\boldsymbol{\pi},\mathbf{x})$, we called one allocation policy $\boldsymbol{\pi}^\ast$ is efficient \textit{iff} for any $\mathbf{t}'$, $\boldsymbol{\pi}^\ast \in \arg \max_{\boldsymbol{\pi}'} SW(\boldsymbol{\pi}',\mathbf{t}')$. Taking single unit auction as an example, efficient allocation policy implies assigning the commodity to the buyer with the highest bid. Secondly, \textit{weakly budget balanced}: one diffusion mechanism $\mathcal{M}=(\boldsymbol{\pi},\mathbf{x})$ is weakly budget balanced if for any $\mathbf{t}'$, $Rev^{\mathcal{M}}(\mathbf{t}')=\sum_{i\in N_{-s}}x_i(t_i',\mathbf{t}_{-i}')\geq 0$ always holds. That is to say the seller $s$ will never incur a deficit.

\subsection{Incentive Compatibility in Diffusion Auction}
 One key feature in auctions is incentive compatibility, an auction mechanism is called incentive-compatible (IC) if every bidder can achieve her best outcome just by acting according to their true type. Another significant concept in auctions is individual rational (IR), which means any bidder will never suffer from a loss as long as she truthfully reports her valuation. The well-known Myerson's Lemma \cite{myerson1981optimal} gave a framework for finding IR and IC mechanisms in a single-parameter environment as follows. 
\begin{table*}
\renewcommand\arraystretch{1.1}
\centering
\scalebox{0.98}{
\begin{tabular}{llll}
\toprule
$\mathcal{M}$ &  Allocation: $\boldsymbol{\pi}$ & Payment: $\mathbf{x}$ & $Rev^{\mathcal{M}}$ \\
\midrule
VCG & $SW^\ast(\mathbf{t}') $ & $SW(\mathbf{t}_{-i}')-(SW(\mathbf{t}')-\pi_i v_i)$ & $\sum_{i\in C_{w}\backslash\{1\}}(v^\ast_{-i}-v^\ast_{N_{-s}})+v^\ast_{-1}$\\ 

IDM     & $SW^\ast(\mathbf{t}_{-(i+1)}')$   & $SW(\mathbf{t}_{-i}')-(SW(\mathbf{t}_{-(i+1)}')-\pi_i v_i)$ &$v^\ast_{-1}$    \\

CDM     & $SW^\ast(\mathbf{t}_{-\alpha_i}')$  & $SW(\mathbf{t}_{-i}')-(SW(\mathbf{t}_{-\alpha_i}')-\pi_i v_i)$   &    $\sum_{i\in C_w\backslash\{1,w\}}(v^\ast_{-(i+1)}-v^\ast_{-\alpha_i}) + v^\ast_{-1}$  \\

FDM     & $SW^\ast(\mathbf{t}_{-(i+1)\cup M_{i,i+1}}')$  & $SW(\mathbf{t}_{-i}')-(SW(\mathbf{t}_{-(i+1)\cup M_{i,i+1}}')-\pi_i v_i)-R_i$ & $R_{s} + v^\ast_{-1}$   \\

NRM     & $SW^\ast({\mathbf{t}'_{-T(i)}})$  &$SW(\mathbf{t}'_{-i})-(SW(\mathbf{t}'_{-T(i)})-\pi_i v_i)-R'_i$& $0 \,(|N_{-s}|\to +\infty)$  \\
\bottomrule
\end{tabular}}
\caption{Table for comparing single-item diffusion auctions. Denote $w$ by the winner satisfying $SW^\ast(\cdot)$ function, $C_w$ is the critical diffusion sequence and only those bidders in $C_w$ have non-zero payment. $\alpha_i$ is an edge set satisfying three properties in CDM. $M_{i,i+1}$ in FDM represents the set of nodes connecting critical node $i$ and $i+1$. $R_i$ and $R_i'$ are different reward functions in FDM and NRM.}
\label{table1}
\end{table*}

\begin{theorem}
(Myerson's Lemma). Fix a single-parameter auction environment: An allocation rule $\boldsymbol{\pi}$ is implementable \textit{iff} it is value-monotonic, and if the allocation rule $\boldsymbol{\pi}$ is value-monotonic, then there is a unique payment rule $\mathbf{x}$ for which $\mathcal{M}=(\boldsymbol{\pi},\mathbf{x})$ is IR and IC, the winner takes the item and pays the critical bid while all other losers' payment is zero.
\end{theorem}

When taking social networks and information diffusion into account, Myerson's Lemma no longer applies because strategic diffusion affects agents' critical bids and payment. Also, the definitions of IR and IC change in diffusion auctions: IR represents no matter how one bidder diffuses the sale information, reporting true valuation ensures her non-negative returns while IC is modified as truthfully reporting her valuation and propagating sale information to all her neighbors will always maximize her utility. 

To formulate IC for diffusion auctions, Li \emph{et al.} \shortcite{li2020incentive} decouples the payment $x_i$ as $x_i(\mathbf{t}')=\pi_i(\mathbf{t}')\tilde{x}_i(\mathbf{t}')+(1-\pi_i(\mathbf{t}'))\bar{x}_i(\mathbf{t}')$. Here $\tilde{x}_i$ represents bidder $i$'s payment for winning the item and $\bar{x}_i$ represents her payment for losing the item which can be intuitively considered as the reward for diffusion. They show that to ensure a diffusion auction to be IC, the allocation rule should be monotone over the bid; the decoupled payments should be independent of the bid; for a fixed $r_i$, the winning payment and losing payment should be monotone over the diffusion effort and the difference between them should equal the bidder's critical bid. More formally, they proved the following theorem.
\begin{theorem}\label{t1}
A single item diffusion auction is dominant-strategy incentive-compatible iff (A) $\boldsymbol{\pi}$ is value-monotonic;
(B) $\tilde{x}_i$ and $\bar{x}_i$ are bid-independent;
(C) $\tilde{x}_i(r_i)- \bar{x}_i(r_i)=v^\ast_i(r_i)$;
(D) $\tilde{x}_i$ and $\bar{x}_i$ are diffusion-monotonic;
(E) $\bar{x}_i(\emptyset) \leq 0$.
\end{theorem}



After identifying the truthful diffusion auctions, they also show that the optimal revenue for seller $s$ under the truthful diffusion auctions can be calculated as:
\[
    Rev(\mathbf{t})=\sum \nolimits_{i\in N_{-s}} \left[  v^\ast_{i}(\emptyset)-v^\ast_i(r_i)(1-\pi_i(\mathbf{t}))\right].
\]
From the perspective of the seller $s$, if bidder $i$ is the winner, her optimal payment is the critical bid when she does not diffuse: $v^\ast_i(\emptyset)$, otherwise, the optimal payment should be the difference between critical bid under no diffusion and truthful diffusion: $v^\ast_i(\emptyset)-v^\ast_i(r_i)$.

\section{Diffusion Auction Mechanisms}
\subsection{Diffusion Auction in Unweighted Network}
The initial work of diffusion auction in \cite{li2017mechanism} 
assumes the diffusion cost is negligible, which can be captured by unweighted networks. They proposed the \emph{information diffusion mechanism} (IDM), which firstly introduced the concepts of critical diffusion nodes and critical diffusion sequences. Then the diffusion process in the social network can be captured by a tree on which the auction is conducted. These concepts and methods are also utilized in the follow-up works of diffusion mechanisms. Consider two bidders $i$ and $j$, $i$ is $j$'s critical diffusion node means $j$ could not participate in this auction without $i$'s diffusion and all of $j$'s critical diffusion nodes form her critical diffusion sequence.

Li \textit{et al.} \shortcite{li2019diffusion} further investigated a class of diffusion auction mechanisms under social networks, which is named as \emph{critical diffusion mechanisms} (CDM). Also, they found that IDM is a special case of this newly discovered set of auction mechanisms, which has the lowest seller's revenue in this set. CDMs rely on the critical diffusion nodes and sequence. However, according to the theorem of small-world \cite{amaral2000classes}, cut-points rarely exist in large well-connected real social network.  
Thus under CDM, the majority of normal nodes may not have strong incentives to diffuse.

 Inspired by the redistribution mechanisms in \cite{cavallo2006Redistribution,guo2009redistribution,guo2012redistribution}, Zhang \emph{et al.} \shortcite{zhangECAIincentivize} proposed the \emph{fair diffusion mechanism} (FDM) which benefit not only cut nodes but also those nodes who have made contributions for the connection from the seller to the winner. Later, Zhang \emph{et al.} \shortcite{zhang2020redistribution} investigated a redistribution mechanism on networks called \emph{network-based redistribution mechanism} (NRM) for more efficient resource allocation. NRM has been proved to be IR, IC and asymptotically budget-balanced.
 Another work \cite{zhang2019fixedprice} in single-unit diffusion auctions without diffusion allies the information diffusion model into fixed-price mechanism design and studied \emph{fixed price diffusion mechanism} (FPDM), which promises the seller $1/2$ optimal revenue.

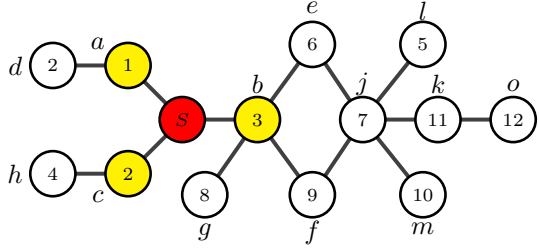
\begin{figure}[!htbp]
    \centering
    \begin{tikzpicture}
\Vertex[x=0,y=0,color=red,label=$S$]{S}
\Vertex[x=1,y=0,color=yellow,label=$3$]{B}
\Text[x=1,y=0.48]{$b$}
\Vertex[x=-0.72,y=0.72,color=yellow,label=$1$]{A}
\Text[x=-1.12,y=1.02]{$a$}
\Vertex[x=-0.72,y=-0.72,color=yellow,label=$2$]{C}
\Text[x=-1.12,y=-1.02]{$c$}
\Vertex[x=-1.72,y=0.72,color=white,label=$2$]{D}
\Text[x=-2.22,y=0.72]{$d$}
\Vertex[x=1.73,y=1,color=white,label=$6$]{E}
\Text[x=1.73,y=1.5]{$e$}
\Vertex[x=1.73,y=-1,color=white,label=$9$]{F}
\Text[x=1.73,y=-1.5]{$f$}
\Vertex[x=0.3,y=-1,color=white,label=$8$]{G}
\Text[x=0.3,y=-1.5]{$g$}
\Vertex[x=-1.72,y=-0.72,color=white,label=$4$]{H}
\Text[x=-2.22,y=-0.72]{$h$}
\Vertex[x=2.4,y=0,color=white,label=$7$]{J}
\Text[x=2.4,y=0.48]{$j$}
\Vertex[x=3.4,y=0,color=white,label=$11$]{K}
\Text[x=3.4,y=0.48]{$k$}
\Vertex[x=4.4,y=0,color=white,label=$12$]{O}
\Text[x=4.4,y=0.45]{$o$}
\Vertex[x=3.2,y=1,color=white,label=$5$]{L}
\Text[x=3.2,y=1.45]{$l$}
\Vertex[x=3.2,y=-1,color=white,label=$10$]{M}
\Text[x=3.2,y=-1.45]{$m$}
\Edge(S)(A)
\Edge(S)(C)
\Edge(S)(B)
\Edge(A)(D)
\Edge(B)(E)
\Edge(B)(F)
\Edge(B)(G)
\Edge(E)(J)
\Edge(F)(J)
\Edge(J)(K)
\Edge(C)(H)
\Edge(J)(L)
\Edge(J)(M)
\Edge(K)(O)
    \end{tikzpicture}
    \caption{Instance of auction in social networks. The colored nodes constitute a local community where the classic auction is conducted. Without information diffusion, the agents with high valuations can not join the sale.}
    \label{fig2}
\end{figure}

Table \ref{table1} summarizes the allocation rules, payment rules and revenues of all single-item diffusion auction mechanisms. We make comparisons of the above diffusion auction mechanisms in the example network in Figure \ref{fig2} and present the results among these mechanisms in Table \ref{table2}. With the social network in Figure \ref{fig2}, initially, the seller $s$ could only sell one commodity among agents $\{a,b,c\}$. Bidder $b$ would win and pay $2$ under a second-price auction. However, when truthful diffusion mechanisms are applied, all participants are willing to propagate the sale information. Thus more potential bidders such as $\{k,m,o\}$ will join. 

Because CDM represents a family of diffusion mechanisms, here we take $\alpha_i$ as the minimum cutting edges subset which makes the next critical diffusion node can not attend the auction as one special case. From Table \ref{table2}, we can see different single unit auction mechanisms have different advantages. VCG guarantees allocation efficiency but not promises the seller's revenue; IDM is a particular case of CDM. The family of these mechanisms sacrifice efficiency but benefit the seller with higher profit. FDM and NRM aim for rewarding not only cut nodes but also those making contributions for information propagation, thus benefiting more reasonable resource allocation in society.

 \begin{table}[!htbp]
    \renewcommand\arraystretch{1.2}
    \centering
    \scalebox{0.9}{
    \begin{tabular}{lllll}
    \toprule
    $\mathcal{M}$ & Winner  & Rewarded bidders & $SW$ &$Rev^{\mathcal{M}}$  \\
    \midrule
    SPA & $b(2)$ & $\emptyset$ & $3$ & $2$ \\
    VCG & $o(11)$  & $b(-8),j(-3),k(-2)$ & $12$  &  $-2$\\
    IDM & $k(10)$  & $b(-5),j(-1)$ & $11$  &$4$\\
    CDM & $k(10)$  & $b(-4),j(-1)$&$11$& $5$\\
    FDM & $k(10)$  & $b(-4),e(-\frac{1}{3}),j(-1)$ &$11$ & $\frac{14}{3}$\\
    NRM & $k(10)$   & $a(-\frac{8}{13}),b(-\frac{18}{13}),c(-\frac{4}{13})$ &$11$& $\frac{61}{26}$\\
     &  & $e(-\frac{5}{8}),f(-\frac{1}{2}),g(-\frac{5}{8})$ & & \\
     & & $j(-\frac{5}{2}),l(-\frac{1}{4})$ & &\\
     \bottomrule
    \end{tabular}}
    \caption{Numerical results for different diffusion auctions run in the network given in Figure \ref{fig2}. Numbers in the parentheses after the nodes are payments. Positive value means the agent pays to the seller while negative value means the seller pays to the agent.}
    \label{table2}
\end{table}

\subsection{Diffusion Auction in Weighted Network}
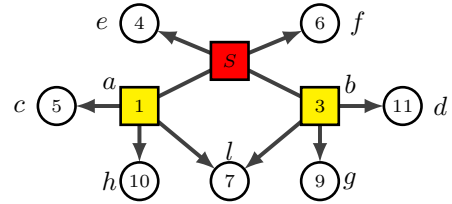
\begin{figure}[!b]
    \centering
    \begin{tikzpicture}
    \SetVertexStyle[Shape=rectangle]
    \Vertex[x=0,y=0,size=0.5,label=$S$,color=red]{S}
    \Vertex[x=-1.2,y=-0.6,size=0.5,label=$1$,color=yellow]{A}
    \Text[x=-1.6,y=-0.3]{$a$}
    \Vertex[x=1.2,y=-0.6,size=0.5,label=$3$,color=yellow]{B}
    \Text[x=1.6,y=-0.3]{$b$}
    \SetVertexStyle[Shape=circle]
    \Vertex[x=-2.3,y=-0.6,size=0.5,label=$5$,color=white]{C}
    \Text[x=-2.8,y=-0.6]{$c$}
    \Vertex[x=2.3,y=-0.6,size=0.5,label=$11$,color=white]{D}
    \Text[x=2.8,y=-0.6]{$d$}
    \Vertex[x=-1.2,y=0.5,size=0.5,label=$4$,color=white]{E}
    \Text[x=-1.7,y=0.5]{$e$}
    \Vertex[x=1.2,y=0.5,size=0.5,label=$6$,color=white]{F}
    \Text[x=1.7,y=0.5]{$f$}
    \Vertex[x=1.2,y=-1.6,size=0.5,label=$9$,color=white]{G}
    \Text[x=1.6,y=-1.6]{$g$}
    \Vertex[x=-1.2,y=-1.6,size=0.5,label=$10$,color=white]{H}
    \Text[x=-1.6,y=-1.6]{$h$}
    \Vertex[x=0,y=-1.6,size=0.5,label=$7$,color=white]{L}
    \Text[x=0,y=-1.2]{$l$}
    \Edge(S)(A)
    \Edge(S)(B)
    \Edge[Direct=True](S)(E)
    \Edge[Direct=True](S)(F)
    \Edge[Direct=True](A)(C)
    \Edge[Direct=True](B)(D)
    \Edge[Direct=True](A)(H)
    \Edge[Direct=True](B)(G)
    \Edge[Direct=True](A)(L)
    \Edge[Direct=True](B)(L)
  \end{tikzpicture}
    \caption{Economic network example: The red rectangle node $s$ is the seller, yellow rectangle nodes $\{a,b\}$ are the intermediaries, and all other circle nodes $\{c,d,e,f,g,h,l\}$ are those potential buyers. Numbers in rectangle nodes represent information diffusion cost.}
    \label{fig3}
\end{figure}

In many cases, we assume the information dissemination is cost-free. However, it is often the scenario that information diffusion in social networks requires fees. Leduc \emph{et al.} \shortcite{leduc2017pricing} focused on referral payments in social networks. Condorrelli \emph{et al.} \shortcite{condorelli2018selling} studied such a setting where the seller sells items to buyers only through intermediaries who can be either merchants or referrals.
An example of weighted economic networks is shown in Figure \ref{fig3}. In such economic networks, all the potential buyers consist of the buyers set $B$; denote $I$ by the intermediaries set and $N=B\cup I\cup \{s\}$ by all agents set. Buyer $i$'s type is defined as $t_i=(v_i,\emptyset)$, where $v_i$ is her valuation for the item and $r_i=\emptyset$ means that she is not able to diffuse the sale information; intermediary $k$'s type is defined as $t_k=(c_k,r_k)$, where $c_k$ is the cost she charges for one successful trading and $r_k$ is the set of her customers.

Li \emph{et al.} \shortcite{li2018customer} is the first to explore auctions in such weighted economic networks. Their \emph{customer sharing mechanism} (CSM) allocates the item to a potential buyer whose trading path would maximize the social welfare. One trading path consists of a seller $s$, a final buyer $i\in B$ and a series of intermediaries $k \in I$. Social welfare in economic networks is different from that in social networks without transfer cost and it is redefined as: $SW=\pi_i v_i - \sum_{k\in I} c_k$. CSM charges the buyer $i$ VCG's payment: the social welfare decrease of the other agents caused by $i$'s participation, and charges the intermediary $k$ the social welfare decrease correlated with their threshold neighbourhood set $r_{k}^{\ast}$. $r_{k}^{\ast}$ represents a minimum subset of $r_k$ where $k$'s diffusion strategy changing from $r_k$ to $r_k\backslash r_{k}^{\ast}$ could change the winner under efficient allocation.

Li \emph{et al.} \shortcite{li2020Access} extended market resource allocation under particular economic networks into a more general setting where intermediaries can not only diffuse sale information to those bidders but also can be potential buyers. Still, all bidders' access to the sale is from the seller and intermediaries' invitation. Their mechanisms \emph{Single-Level Diffusion Mechanism} (SLDM) and \emph{Multi-Level Diffusion Mechanism} (MLDM) applies to the single level distribution market scenarios where all intermediaries are connected to the seller and multiple-level markets where intermediaries diffuse the information in a tree structure respectively. 

In the above economic networks, information flows in one direction from the seller through intermediaries to buyers, which means those buyers cannot diffuse sale information. They can never participate in these auctions without the seller and the intermediaries. Li \emph{et al.} \shortcite{li2019diffusion} generalized economic networks to a weighted graph where all participants could bid and propagate sale information and weights of edges represent transfer cost between agents. Further, they illustrated the differences between auctions via unweighted networks and weighted ones, explained why all previous mechanisms cannot be applied to a weighted graph scenario and proposed \emph{weighted diffusion mechanism} (WDM), which can deal with this challenging problem. 
They illustrated that the minimum cost allocation path conflicts with the critical propagation path in weighted networks. WDM allocates along the minimum cost trading path and assigns the item to the first bidder $i$ whose bid is highest when her diffusion strategy changes $r_i$ to $r_i\backslash \gamma^\ast_i$. Here $\gamma^\ast_i$ is a special edge set containing $i$ and the concatenated edges of those non-terminal nodes in $r_i$. All agents in minimum cost trading path pay $SW^\ast(t_{-i}')-SW^\ast(t_{-\gamma^\ast_i}')$ and the winner undertakes the critical bid that could beat all nodes before her in the trading path with minimum cost.

Comparing diffusion auctions in weighted networks and unweighted networks, without diffusion cost, we focus on the critical diffusion sequences and allocate the item to the winner in any trading path. However, when considering cost, it changes a lot since the cost is related to efficiency, bidders' allocation and seller's revenue, thus all these with-cost diffusion mechanisms focus more on the trading path with minimum cost rather than the critical diffusion sequence in cost-free diffusion auctions. This is the most significant difference between these two scenarios.

 \subsection{Non-Truthful Diffusion Auction}
 From the well-known work \cite{29088}, we can know that efficiency and weakly budget-balance for designing mechanisms in general quasi-linear domains cannot be achieved at the same time. This is also true in diffusion auctions. Some diffusion mechanisms sacrifice efficiency to ensure weakly budget-balance while some others maintain efficiency with a deficit, such as VCG
 with diffusion. However, there exist another branch of diffusion auctions where IC is sacrificed, ensuring efficiency and weakly budget-balance.

The very first work in this domain is from MLM\cite{lee2017mechanisms}. Based on the fact that VCG diffusion mechanism is efficient, IC, IR but not budget feasible, they proposed their \textit{multilevel mechanisms} (MLM). In a single-item diffusion auction, MLM allocates the item to the highest bidder. Its payment rule makes some improvement on the VCG payment, from individual externality to groupwise externality. The specific differences between individual externality ($x^{\text{ie}}_i$) and groupwise externality ($x^{\text{ge}}_i$) are as follows:
\[
    \begin{split}
        x_i^{\text{ie}} &= SW^\ast(\mathbf{t}_{-i}')-(SW^\ast(\mathbf{t}')-\pi_iv_i),\\
        x_i^{\text{ge}} &= SW^\ast(\mathbf{t}_{-i}')-(SW^\ast(\mathbf{t}')-\sum \nolimits_{j\in T(i)}\pi_j v_j).
    \end{split}
\]
Note $T(i)$ represents the set of bidders under $i$'s critical diffusion sub-tree. Further, the payment rules of VCG equals $x_i^{\text{ie}}$ and MLM can be illustrated as follows:
\[
    x_i^{\text{MLM}} = x_i^{\text{ge}} - \sum \nolimits_{k\in C(i)}  x_k^{\text{ge}}.
\]
Note that $C(i)$ represents the set of $i$'s immediate children, specifically, all agents who are $i$'s first-order neighbors while $i$ is their critical diffusion node.

MLM is efficient, weakly budget balanced, IR but not IC. Buyers could manipulate the auction by misreporting higher bids. Some analysis and improvements on the MLM have been made in \cite{jeong2020referrer} and \cite{jeong2020groupwise}. Jeong \shortcite{jeong2020referrer} introduced referral monotonicity: when all other bidders' types are fixed, the utility will never decrease with her diffusion strategy for every potential buyer, then they proposed the \textit{Referrer's Dilemma}: although more diffusion brings non-decrease profit from individuals, when all bidders try to propagate information, they are weakly worse off but the seller weakly increases her revenue from a global perspective. They examined which mechanisms are subject to this dilemma and the impact of false-name attacks and collisions on these mechanisms. Jeong and Lee \shortcite{jeong2020groupwise} renamed the \textit{multilevel mechanism} as \textit{groupwise-pivotal referral mechanism} (GPR) and proved it to be groupwise collusion-proof (any complicity in groups can not improve utility) and Sybil-proof (preventing false-name manipulation).


 
\subsection{Multi-Unit Diffusion Auction}
A classic multi-unit auction allocates multiple homogeneous items to bidders. Multi-unit auctions are widely used in different social sectors for electromagnetic spectrum\cite{cramton1997fcc,cramton2000collusive}, electricity distribution, government securities, etc. When extending multi-unit auctions into social networks, it becomes a very complicated decision-making problem since ancestor bidders can control items passed to their children, influencing their payments. At the same time, they can change their peers' payments without changing their allocation and payments. 

Even in multi-unit auctions with single-unit demand where many results \cite{krishna2009auction} in single-item auctions can be extended, it becomes quite complex in the context of diffusion. Zhao \emph{et al.} \shortcite{zhao2019sellmultipleitems} firstly proposed the \emph{general information diffusion mechanism} (GIDM) to solve multi-unit diffusion auctions with single-unit demand. They used a critical diffusion tree based on the critical diffusion sequence in IDM and divided it into some sub-markets. Considering one $k$-unit diffusion auction with single-unit demand, all the top-$k$ bidders' critical diffusion sequences consist of the critical diffusion tree and whether top-$k$ bidders can get one item or not depends on their ancestors. If one ancestor is eligible for an item, then the original top-$k$ node with the lowest bid under the ancestor's sub-tree will cede the item. GIDM is the first attempt to tackle the multi-unit diffusion auction design and it inspires a few follow-up works on this complicated task. 

Takanashi \emph{et al.} \shortcite{takanashi2019efficiency} studied \emph{Generalized Aligned Path Graph Mechanism} (GAPG) for multi-unit auctions with decreasing marginal utility, which is IC, IR but not weakly budget balanced.
Further, \emph{Distance-based Network Auction Mechanism} (DNA-MU) for multi-unit diffusion auction with single-item demand was proposed in \cite{kawasaki2020strategy}. It allocates items in the order of distance from the seller to buyers and one buyer $i$ is qualified to win an item when her reported price satisfies $v_i' \geq v^k(\mathbf{t}'_{-\{i\}\cup W})$. Here $W$ is the winner set and once a buyer is assigned one item, she will be added into $W$. The payment of winner $i$ is this critical bid $v^k(\mathbf{t}'_{-\{i\}\cup W})$. Whenever an item is assigned out, the total number of items $k$ is self-reduced by $1$. The whole process finishes when all $k$ items are allocated. DNA-MU mechanism is proved to be IR and weakly IC, which means agents have no direct incentives to invite their friends to join since they can not directly gain reward from diffusion. More detailed analysis of efficiency and revenue are finely illustrated in this work. It gave a good viewpoint that the quantified network structure criteria such as path length, degree, clustering coefficient can be utilized for the diffusion auction design.

Condorrelli \emph{et al.} \shortcite{condorelli2017bilateral} studied bilateral trading in networks. Xu \emph{et al.} \shortcite{xu2019double} introduced networked double auction with multiple sellers and buyers existing in their problem setting. Their mechanism \emph{Double Network Auction Mechanism} (DNAM) combines McAfee's mechanism \cite{mcafee1993mechanism} with VCG mechanism. DNAM can be degenerated into single-unit cases since their sub-markets are independent.

Researches on multi-unit diffusion auctions are incomprehensive yet and there still remains many challenging problems. For instance, monotonic allocation rules involving diffusion is difficult to identify, how to restrict the bidders' action space to simplify the outcome space is still unclear, etc.


\section{Diffusion Incentive Design beyond Auctions}
There are several application directions where incentive matters but there is no bid for purchasing. Such applications include crowdsourcing, matching, voting and so on. 


Crowdsourcing \cite{howe2006rise} is a collaboration model where some institutions try to attract potential employees to solve a colossal task, which takes the form of peer-production. All the agents involved solve the job collaboratively (e.g. LEGO Ideas, Amazon Mechanical Turk). In traditional crowdsourcing problems, the requester pays not only employees but also third-party platforms with a high cost; the quality of sourcing from the paid platforms may not be guaranteed. However, employees' recruitment can be done spontaneously when considering the collaboration via social networks. Since employees are rewarded directly from the requester, the quality and flexibility are also improved.

DARPA 2009 red balloon challenge is a typical example in this field. This challenge awards the first team who submits positions of all ten weather red balloons in the continental United States. Pickard \emph{et al.} \shortcite{pickard2011time} gave the winner solution for this task: \textit{recursive incentive mechanism} (RIM), which diffuses task information through social networks and provides incentives for individuals who give solutions and make recruitment. The reward takes a bottom-up cascading form. Cebrain \emph{et al.} \shortcite{cebrian2012finding} provided a split contract mechanism in the same setting. Each individual receives rewards from her parents and decides the reward offer to her children nodes in their mechanism. Emek \emph{et al.} \shortcite{emek2011mechanisms} deals with multi-level marketing mechanisms in which small rewards are allocated for preventing false-name manipulations. Zhang \emph{et al.} \shortcite{zhang2020sybil} studied a question answering problem. 
They characterized the incompatibility between Sybil-proofness and collusion-proofness under strong IR assumption and proposed \textit{Double Geometric Mechanism} (DGM), which is Sybil-proof and $2-$collusion-proof. 

Zhang \emph{et al.} \shortcite{zhang2019collaborative} focused on another crowdsourcing instance called data acquisition, where a requester acquires useful data (e.g. images, corpus) via social networks. Their target is to incentivize all agents to provide all their valid data and inviting their neighbors. their novel \textit{crowdsourcing diffusion mechanism} rewards one agent for her contributions to data and diffusion. \textit{Shapley value} describes how to distribute both gains and costs to several agents working in a coalition\cite{roth1988shapley}. They modified it into the layered Shapley value since the traditional form discourages agents from inviting their friends. Agents with the same depth in the diffusion network are treated as one group and their reward equals the marginal contribution for their participation. Also, information entropy\cite{shannon1948mathematical} is used as a valuation function in layered Shapley value.

Zhang \emph{et al.} \shortcite{zhang2020coalitions} extended information diffusion settings into a cooperative game. To incentivize agents' diffusion, they combined weighted Shapley value\cite{RADZIK2012407} which is applied in cooperative asymmetry game, with permission structure\cite{gilles1992games} which requires that some agents could join in the coalition only if getting permission from other agents. Shi \emph{et al.} \shortcite{shi2019maximalpropagate} focused on an information diffusion game via social networks where one requester wants to spread information via social networks, paying the participants for their diffusion with a fixed budget.
Their scheme contains two stages: the first is hierarchizing the entire diffusion tree and deciding the sum of rewards for each layer based on the total budget while the second is determining rewards for agents in each layer respectively.

Social choice problems like matching, voting, rating under network scenarios are also important research fields. The pioneer work in this thread is from \cite{10.5555/2772879.2773278}, which is later generalized in \cite{grandi2017strategic,grandi2020personalised}. They studied the opinion diffusion problem with multiple agents in social network settings and showed how one agent is affected by the opinions of those agents that she trust, like relatives, neighbors or good friends. The social network structure is a key component in analyzing and modeling the trust and reputation. 

Some other works \cite{zheng2020barter,2021networkedhousingmarket} focus on the networked house allocation problem. Kawasaki \emph{et al.} \shortcite{2021networkedhousingmarket} defined new networked core concepts and illustrated incompatibility between networked core and strategyproofness. They restricted preference into acyclic domain to guarantee stratgyproofness of TTC in networked market and  proposed modified TTC under tree networked market.


\section{Discussion}

Past few years have witnessed considerable progress in this emerging networked auction. Generally, we have highlighted single item networked auction  design and multi-unit with single demand, diffusion mechanisms focus on incentivizing truthful bidding and inviting neighbors to join, requiring tradeoff between different properties. However, tasks about multi-unit and combinatorial auctions via social networks have not yet 
been explored clearly. It is also a valuable topic to integrate network structures into property analysis in diffusion mechanisms design. Besides, Sybil-proofness \cite{todo2020split} and collusion-proofness in networked auction are both nice directions. 

There are other representative scenarios where the incentive design of information diffusion deserves further investigation beyond auction. For example, how to conduct a more credible election or voting \cite{ColleyEtAlIJCAI2020} by encouraging more people to join; how to facilitate a good organ donation \cite{roth2004kidney,roth2005pairwise} and transplantation system in a networked scenario; how to incentivize more people into a rating system; and how to notice more citizens to conduct public opinion surveys. Any other mechanism design scenarios where attracting more participants is needed can also be extended into social network settings.


\bibliographystyle{named}
\bibliography{diffusion}

\end{document}